# Magnetically collected platinum/nickel alloy nanoparticles – insight into low noble metal content catalysts for hydrogen evolution reaction

*Sebastian Ekeroth[a], Joakim Ekspong[b], Sachin Sharma[a], Robert Boyd[a], Nils Brenning[a, c], Eduardo Gracia-Espino[b], Ludvig Edman[b], Ulf Helmersson[a]*, Thomas Wågberg[b]**

[a)] Department of Physics, Chemistry and Biology, Linköping University, SE-581 83 Linköping, Sweden.

[b)] Department of Physics, Umeå University, SE-901 87 Umeå, Sweden.

[c)] KTH Royal Institute of Technology, School of Electrical Engineering, Division of Space and Plasma Physics, SE-100 44 Stockholm, Sweden.

E-mail: Thomas.wagberg@umu.se

## Abstract

The hydrogen evolution reaction (HER) is a key process in electrochemical water splitting. To lower the cost and environmental impact of this process, it is highly motivated to develop electrocatalysts with low or no content of noble metals. Here we report on a novel and ingenious synthesis of hybrid $Pt_xNi_{1-x}$ electrocatalysts in the form of a nanoparticle-necklace structure named nanotrusses, with very low noble metal content. The nanotruss structure possesses important features, such as good conductivity, high surface area, strong interlinking and substrate adhesion, which renders for an excellent HER activity. Specifically, the best performing $Pt_{0.05}Ni_{0.95}$ sample, demonstrates a Tafel slope of 30 mV dec$^{-1}$ in 0.5 M $H_2SO_4$, and an overpotential of 20 mV at a current density of 10 mA cm$^{-2}$ with high stability. The impressive catalytic performance is further rationalized in a theoretical study, which provides insight into the mechanism for how such small platinum content can allow for close-to-optimal adsorption energies for hydrogen.

## 1. Introduction

Hydrogen is a storable energy carrier that can be transformed to electric energy in hydrogen fuel cells. If the hydrogen is produced electrochemically by electrolysis using renewable energy as the driving source, it would represent a sustainable fuel, in particular if the two electrochemical half reactions can proceed with low energy losses. The electrocatalysts at the hydrogen evolution cathode and the oxygen evolution anode are the keys to mitigate such undesired energy losses. For the hydrogen evolution reaction (HER) under acidic conditions, Pt has for long been regarded as the champion material since it displays optimal adsorption energies for hydrogen.[1, 2] Although non-noble metal alternatives, such as various transition metals in combination with sulfides,[3-7] selenides,[8, 9] phosphides,[10-12] carbides[13, 14] and nitrides,[15] are reported to exhibit good catalytic activity, they have in most cases not been able to outcompete Pt-based catalysts in regard to both efficiency and stability. Common strategies to lower the requirement on noble metals are to decrease the size of the noble metal nanoparticles,[16-18] and thereby increase the surface/volume ratio, or to alloy the noble metal with transition metals, such as Ni,[19, 20] Co,[21, 22] or Fe.[23, 24] Due to their large differences in electronegativity it is difficult to obtain stable PtNi alloys with controlled Pt:Ni ratio by wet-chemical co-reduction synthesis.[25] Here, we report on the synthesis of a new hybrid material named nanotrusses and its utilization as HER electrocatalysts. The hybrid nanotruss material comprises necklace-like interconnected nanoparticles, with distinct metal-metal interconnections. This leads to very good electrical connection along the whole truss as well as a high surface area due to the interlinked nanoparticles. The catalyst material is produced in a pulsed plasma where the constituent materials form alloyed nanoparticles, which are then collected magnetically into nanotrusses. Earlier studies explained how to collect alloys of two magnetic materials Fe and Ni.[26] Here we show that also magnetic and non-magnetic elements can be intermixed into homogeneous alloys and that their atomic ratio can be precisely tuned by the pulsed plasma parameters. Through this innovative and versatile process, we synthesize PtNi alloys that perform excellent as a HER electrocatalyst with extremely low overpotential along with high stability, comparable or even better than similar Pt-based catalysts.[27-29]

## 2. Experimental section

*2.1. Material synthesis*: The pulsed plasma technique is described in detail in a previous publication.[26] In short, the technique is accomplished by the following procedure. Hollow cathodes of Ni (99.99 %) and Pt (99.95 %) are placed with their openings close to each other. The cathodes, with a length of 54 mm and an inner diameter of 5 mm, are electrically insulated from each other in order for separate electrical pulses to be delivered to each one. The grounded stainless-steel anode ring, 30 mm in diameter, is placed 25 mm from the openings of the cathodes. To confine the plasma an electrically floating cylindrical stainless-steel mesh cage encloses the region defined by the cathode openings and the anode ring. Ar (99.9997 %) is used as the process gas and is flown through the two cathodes at a total rate of 120 sccm. Small amounts of $O_2$ is added, from outside the mesh cage, to support nanoparticle nucleation.[30, 31] To efficiently add and control the $O_2$, it is diluted in Ar (95 %), resulting in an effective flow of 0.025 sccm.

Two HiPSTER 1 pulsing units (Ionautics AB), fed by two MDX-1K dc power supplies (Advanced Energy Industries, Inc), create pulsed plasmas in each of the cathodes. The reason for using high-power pulsed technique is to increase the degree of ionization of the sputtered growth material. By using ions, a fast growth of the nanoparticles is achieved, as the positively charged ions are attracted to the nanoparticles, which obtain a negative charge in the plasma.[32, 33] The pulse sequence is controlled by a HiPSTER Sync unit. The unit is set to deliver 80 μs long pulses, in sequence to the two cathodes, at a total frequency of 1200 Hz, giving an effective pulsing frequency of 600 Hz to each cathode. This pulse pattern is referred to as *continuous pulsing* in earlier work.[34] The power to the individual cathodes is adjusted, by tweaking the pulse voltage, to tune the amount of sputtered material from each cathode. For most process conditions, the sum of the average power, from the two cathodes, is kept at 60 W. This is done to avoid overheating of the cathodes. However, as the Pt cathode does not provide a stable plasma below 3 W, the total power is raised to 100 W for one experimental condition, in order to achieve power ratios below 5 % for the Pt. In order to compensate for the higher average power, the total deposition time is decreased, from 5 minutes to 3 minutes. All of these parameters are presented in the process conditions table (Supporting information, Table S1). The substrates are placed on top of a magnet in order to collect the nanoparticles magnetically, as explained in earlier work.[34-36] However, unlike the previous works, the magnet used here is a permanent cube magnet made from NbFeB, with a 7 mm side.[37] This makes the magnetic flux density at the center of the magnet significantly stronger compared to the earlier work.[34-36] This results in a more efficient collection of the nanoparticles. Si wafers, coated with 200 nm of Ti to yield an electrically conductive surface, are used as substrates for samples used in materials characterization. For the catalysis, depositions are performed onto a carbon-paper substrate. The substrates are attached to the magnet using double adhesive copper tape. The substrate-magnet assembly is placed on the inside of the mesh, at the same lateral distance from the cathode openings as the anode ring and in a spot of the circumference that gives equidistance to both cathodes.

Samples that are electrically collected are placed on a rotatable substrate holder approximately 20 mm outside an opening in the mesh. The holder, that fits 6 samples, sits at approximately 150 mm from the cathode openings facing the cathodes through the anode. For details of the geometrical setup, see reference.[38]

*2.2. Material Characterization*: Grazing incidence X-ray diffraction (GIXRD) experiments are performed using an Empyrean diffractometer in a parallel beam configuration with a line-focused Cu-anode source (Cu $K\alpha_1$ = 1.54 nm), operated at 45 kV and 40 mA. The primary beam is adjusted using a parallel beam mirror and a 1/4° divergence slit and in the secondary beam path a 0.27° parallel plate collimator is set. A PIXcel-3D detector is used for the data acquisition. The grazing incidence X-ray diffraction scans are performed at an incidence angle of 1° in a $2\theta$-range of 30°-80° using a step size of 0.015° and a data acquisition time of 880 ms per step. Scanning electron microscopy (SEM) images were taken using a LEO 1550 instrument with a 5 kV electron beam. All micrographs were taken with a sample tilt of 54 deg. To better visualize the complicated structure of the nanotrusses, two images are recorded with different detectors, one located to the side (SE2) and the other from the top (In-lens), to produce a more informative composite image. Scanning transmission electron microscopy (STEM) and transmission electron microscopy (TEM) analyses were performed on selected samples to determine both the internal nanoparticles structure and the interparticle interactions. Prior to the STEM and TEM analysis, the nanoparticles were placed onto a copper TEM grid coated with an amorphous layer of carbon. Electrically collected particles were deposited directly onto the grids while magnetically deposited particles were transferred from the substrate. All measurements were taken using a FEI Tecnai G2 operated at 200 kV. High angle angular dark field (HAADF) images were taken with a detector spanning an angular range from 80 to 260 mrad. Energy dispersive X-ray spectroscopy (EDS) maps were taken with no sample tilting, an acquisition time of 750 ms per spectra, and a resolution of 2 nm. All XPS data were collected on a Kratos Axis Ultra DLD instrument from Kratos Analytical (UK) employing monochromatic Al Kα radiation (hν = 1486.6 eV). The spectra were collected from a $0.3 \times 0.7$ mm$^2$ area and with electrons emitted along the surface normal. XPS compositional analysis gave a Pt/Ni ratio of 0.06 along with the high concentration of oxygen. The core spectra of Ni 2p and Pt 4f were peak fitted using the peak position provided in [39] and [40], respectively.

*2.3. Electrochemical measurements*: Electrochemical measurements were done in $H_2SO_4$ (15 ml, 0.5 M) with a potentiostat (PGSTAT302N, FRA32M module) connected to a three-electrode setup with Ag/AgCl (CHI111-CH instruments, 1 M KCl) as the reference electrode, a Pt coil as the counter electrode and the coated carbon paper (Sigracet GDL 34 AA) as the working electrode. A circle with geometrical area of 0.2 $cm^2$ was punched out from the coated carbon paper, which was then connected to the potentiostat with a copper wire and sealed with epoxy resin on a microscopic glass slide with only the front side exposed.[41, 42] For the linear sweep voltammetry, the scan rate was 5 mV $s^{-1}$. All potentials were corrected post measurement for the solution resistance, measured to 2.5 Ohm by electrochemical impedance spectroscopy, and presented vs. the reversible hydrogen electrode (RHE) potential by calibrating against a RHE (Gaskatel, Hydroflex). The stability tests were executed with chronoamperometry with a current density of 10 mA $cm^{-2}$ for 5 hours.

*2.4. Computational details*: *Ab initio* calculations of the hydrogen adsorption energies on the catalyst particles were performed using density functional theory with the SIESTA code.[43] The generalized gradient approximation with the revised Perdew, Burke, and Ernzerhof (RPBE) parametrization were used to describe the exchange and correlation functional.[44] The valence electrons were represented using linear combinations of pseudo-atomic numerical orbitals with a double-ζ polarized basis.[45] A mesh cutoff of 250 Ry was used to define the real-space grid. The geometrical optimizations of the metal slabs were carried out using a Monkhorst-Pack k-grid of 3×2×1 and a variable cell scheme by conjugate gradient minimization until the maximum forces were < 0.05 eV $Å^{-1}$.[46] To calculate the hydrogen adsorption energies on the catalyst particles, (111 and 100) slabs of Ni, Pt and NiPt alloys models were constructed, where the slabs consist of four atomic layers. A vacuum of 15 Å was added above the structures in the z direction to avoid interaction between the cells. The Pt atoms were positioned homogenously dispersed in each layer and the alloy was relaxed together with the cell parameters. For the hydrogen adsorption optimizations, the cell parameters and the two bottom layers of the slabs were kept fixed. The hydrogen adsorption energies were calculated using the computational hydrogen electrode model that was further developed for heterogeneous materials, also taking equilibrium hydrogen occupations into account.[47, 48] In this model, the hydrogen occupation is first determined for each atomic site and the free energy of adsorption values are thereafter calculated for the equilibrium structure. The theoretical voltammetry plots along with exchange current densities were then calculated including all adsorption sites following the microkinetic model that was developed for defective and heterogenous materials.[48] The $NiPt_{111}$ and $NiPt_{100}$ phases were evaluated separately and combined without interaction. The formation energies of surface slabs were calculated with the relation $E_{\mathrm{form}} = E_{\mathrm{total}} - N_{\mathrm{Ni}}\mu_{\mathrm{Ni}} - N_{\mathrm{Pt}}\mu_{\mathrm{Pt}}$, where $E_{\mathrm{total}}$ is the total energy of the metal slabs (PtNi alloy, Ni or Pt) and $\mu_{\mathrm{Ni}}$ and $\mu_{\mathrm{Pt}}$ are the chemical potential of Ni and Pt species, respectively, calculated from reference states. $N_{\mathrm{Ni}}$ and $N_{\mathrm{Pt}}$ are the number of Ni and Pt atoms, respectively, in the structure. The adsorption free energy of hydrogen is calculated from the relation $\Delta G_{\mathrm{H}} = \Delta E_{\mathrm{H}} + \Delta E_{\mathrm{ZPE}} - T\Delta S$, where $\Delta E_{\mathrm{H}}$ is the adsorption energy of hydrogen, $\Delta E_{\mathrm{ZPE}}$ and $\Delta S$ are the difference in zero-point energy and entropy of adsorbed hydrogen and hydrogen in gas phase. The latter term $\Delta E_{\mathrm{ZPE}} - T\Delta S$ has been calculated for similar metals and adsorption sites before and is used in our calculations.[49]

For the simulations of the XRD and the segregated Pt-nanoparticles a spherical nanoparticle with a diameter of 20 nm was constructed directly from the fcc lattice of Ni, the resulting particle had a total of 383913 atoms. The Pt doping was carried out in such ways that the top 4 layers (54596 atoms) had an average concentration of 7.3 at%, however 50% of the surface had no Pt, so the other half achieved a concentration of ~15 at%. The remaining Pt atoms were then homogeneously distributed in the rest of the nanoparticle. These conditions are chosen to reflect the experimental conditions, as well as earlier observations on similar type of alloy nanoparticles. [50] Table S2 shows the Pt-distribution in the different nanoparticles following these requirements. The nanoparticles were then geometrically optimized using the large-scale atomic/molecular massively parallel simulator (LAMMPS),[51] and the embedded atom (EAM) potentials for Pt and Ni from B-Adams, et al.[52] No periodic boundary conditions were applied, and only conjugate gradient minimization was performed. The as-optimized structure was then used to simulate the powder X-ray diffraction using Debye scattering formula,[53] as implemented in Debyer code.[54]

## 3. Results and discussion

To achieve the different element compositions of the $Pt_xNi_{1-x}$, $0 \leq x \leq 1$, the power to the two cathodes (Pt and Ni) is regulated separately, yielding different amounts of the sputtered materials in the growth zone of the nanoparticles. The power on the cathodes and other process parameters of the different compositions are given in Table S1, supporting information. **Figure 1**(a) shows grazing incidence X-ray diffraction (GIXRD) of the $Pt_xNi_{1-x}$ samples, which were collected either electrically (red curves) or magnetically (black curves). The full range GIXRD data are shown in Figure S1 and Table S3 in the supporting information.

The composition factor x used throughout the article relates to the nominal composition given by the power ratio supplied to the two cathodes. A verification for that this is a good approximation is seen in Figure 1(b), where the nominal composition is plotted against the calculated composition from both XRD and STEM-EDS measurements (as described further below).

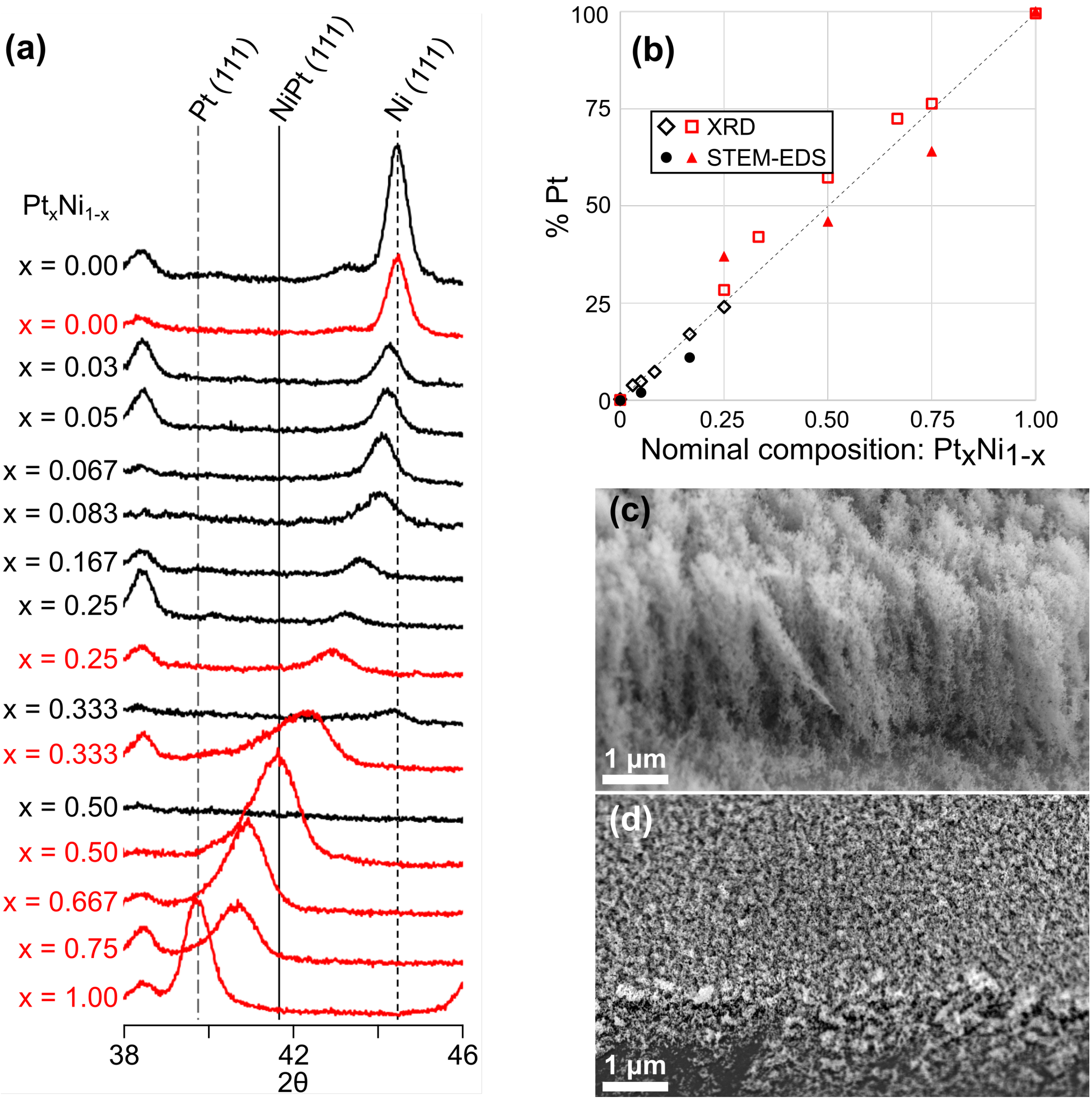

**Figure 1.** Basic characterization of the $Pt_xNi_{1-x}$ samples. (a) Grazing incidence X-ray diffraction of $Pt_xNi_{1-x}$ samples. Magnetically collected nanoparticles are depicted in black, and electrically collected are depicted in red. The nominal $Pt_xNi_{1-x}$ composition is presented to the left of the data. (b) The calculated composition is based on the XRD peak positions in (a) and calculated through Vegard's law and from STEM-EDS measurements. The nominal composition is based on the applied power ratio to the two cathodes. The dashed line is marking the linear 1:1 ratio. Scanning electron microscopy side-view images of (c) magnetically collected nanoparticles with a composition of $Pt_{0.05}Ni_{0.95}$ and (d) electrically collected particles with a composition of $Pt_{0.25}Ni_{0.75}$. The number of particles collected electrically is much smaller than that collected magnetically.

As earlier work has shown, the possibility to collect nanoparticles using the magnetic collection technique requires sufficient magnetic properties of the nanoparticles.[35] To determine the limit for magnetic collection of the $Pt_xNi_{1-x}$ nanoparticles, XRD investigations were used. Figure 1(a) depicts that for magnetically collected samples (black curves), no peaks can be seen for x = 0.50. Moreover, for x = 0.33, we note a small peak corresponding to pure Ni, which is in contrast to samples with lower x that are alloys, as seen in the shift of the Ni (111) peak. As the electrically collected nanoparticles do not form nanotrusses on the substrate, as shown in Figure 1(c-d), it is only the magnetic collection that yields nanotrusses that are desired for the electrocatalytic application. Hence, the composition range for electrocatalytic evaluation is limited to $0 \leq x \leq 0.25$. However, to illustrate the applicability of the deposition technique, the deposition and growth of the nanoparticles are performed over the full range $0 \leq x \leq 1$, and for the higher x-values the nanoparticles had to be electrically collected. The nanotruss structure shown in Figure 1(c) was found mechanically stable and previous studies on the formation of Fe nanotrusses do show, in fact, that the interface between such particles do consist of adhere crystalline planes. Furthermore, the nanotruss structure cannot be easily destroyed, even after treatment with ultrasound significant portion of the structures remain intact. This indicates that there is strong and specific adhesion between the individual particles.

The XRD data for all samples (except the magnetically collected x = 0.33, 0.50) show a dominating peak originating from (111) planes of the crystal. The position of the (111) shifts gradually from Ni(111) to Pt(111) as the nominal composition shifts from pure Ni to pure Pt. In Figure 1(b) the peak position and the shift are used in combination with Vegard's law (assuming an ideal solid solution between Pt and Ni) to estimate the alloy composition through the following relation between the lattice parameters (*a*) of the alloy and the pure elements: $a_{\mathrm{Pt_xNi_{1-x}}} = (1-\mathrm{x})a_{\mathrm{Ni}} + \mathrm{x}a_{\mathrm{Pt}}$.[55] As a complementary method, STEM-EDS was also used to directly analyze individual particles, in the case of electrical collection, or small fragments of nanochains in the case of magnetic collection. Both XRD and STEM-EDS proves that the produced particles are Ni/Pt alloys, rather than a mixture of separate Ni and Pt particles, with the expected compositions. It is worth noting that the number of particles analyzed for STEM-EDS is several orders of magnitude lower that for XRD. For the $Pt_{0.05}Ni_{0.95}$ sample the measured Pt content by STEM-EDS is significantly lower than expected (2 % as opposed to 5 %). This is most likely due to the low signal from the limited amount of Pt and number of particles analyzed leading to measurement error. Additional XPS analysis of this sample gave the Pt concentration of 6 %, slightly larger than expected. XPS, distinct from the other techniques used here, does not probe the entire particle rather only the top 5-8 nm. Pt enrichment at or near the surface could explain its higher recorded content.

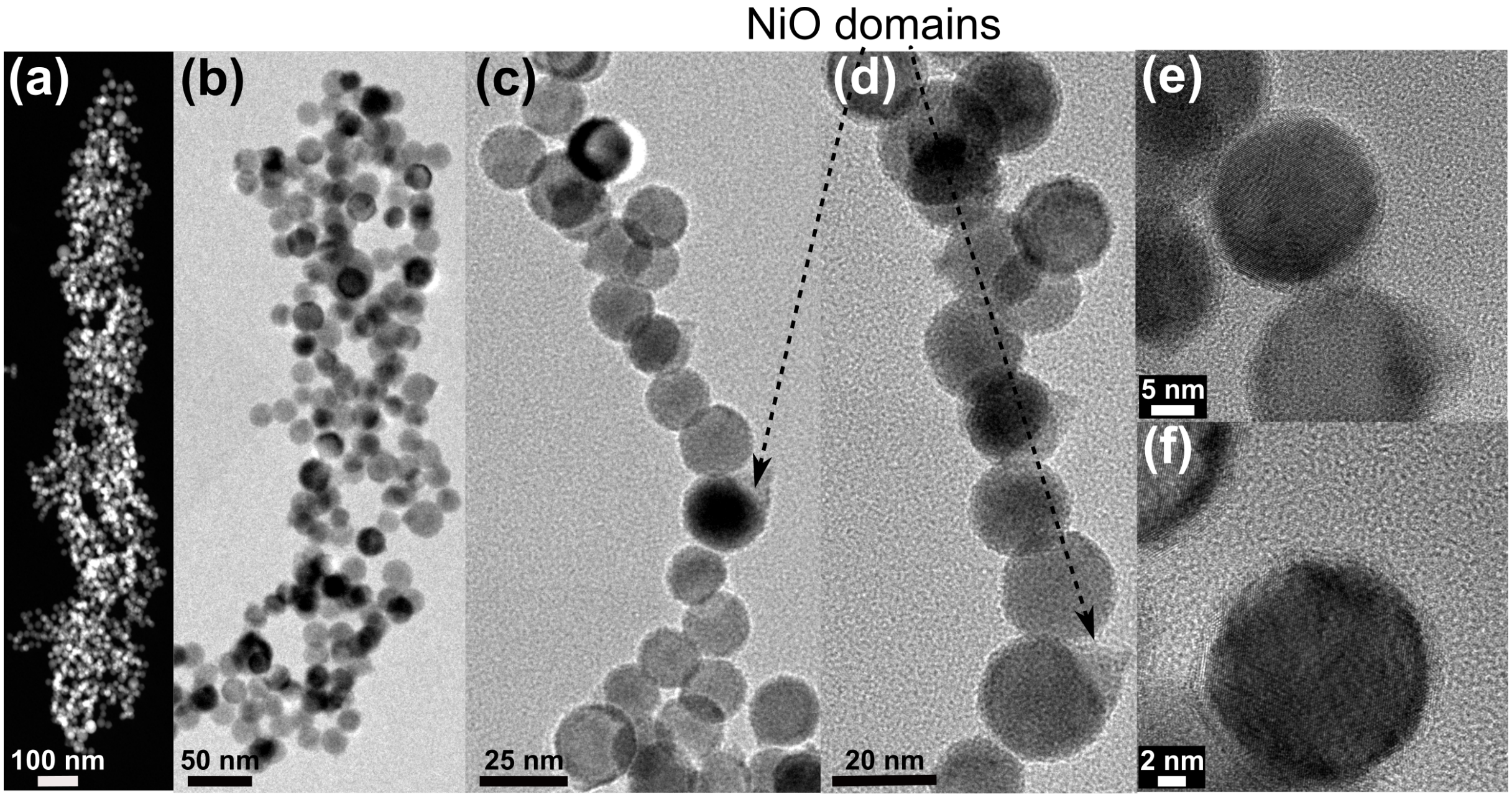

**Figure 2.** STEM highlighting the structure of the $Pt_{0.05}Ni_{0.95}$ nanotruss. (a) HAADF STEM image showing an individual truss. (b-d) bright field TEM images showing individual nanoparticles and the bead-on-a-string structure. The triangular features indicated are NiO domains that have grown on the Ni particles. (e-f) HRTEM images of individual particles showing a single crystal core and an oxide shell.

**Figure 2** shows STEM and HRTEM images of a $Pt_{0.05}Ni_{0.95}$ nanotruss. The overall structure of the nanotruss, Figure 2(a), is highly porous contributing to a high surface-to-volume ratio. Figure 2(b-d) reveals that the nanoparticles self-organize into interlinked nanowires with a beads-on-a-string architecture. The individual nanoparticles are relatively uniform with a size range of 15 nm to 25 nm, and predominately spherically shaped. The only non-spherical deviation is particles featuring protruding triangular features, which have been identified in the past as NiO domains.[35] The HRTEM images in Figure 2(e-f) show that the particles consist of a metallic core surrounded by a thin oxide shell. The XPS data in Figure S2(a-c) provides further support for this conclusion, and after fitting the Ni 2p[39] and Pt 4f peaks,[40] we find that the vast majority of the detectable Ni (97 %) is oxidised whereas 45 % of the Pt is oxidized.

The presence of an oxide shell can have a strong effect on the properties of the particles, and from the HRTEM images in **Figure 3** it appears as though the oxide layers are most pronounced for particles with high Ni content ($Pt_{0.17}Ni_{0.83}$ and pure Ni). The data reveals that the vast majority of the particles consist of a single crystal domain with a small minority consisting of 2 or 3 distinct crystal domains, with no apparent separation into different phases.

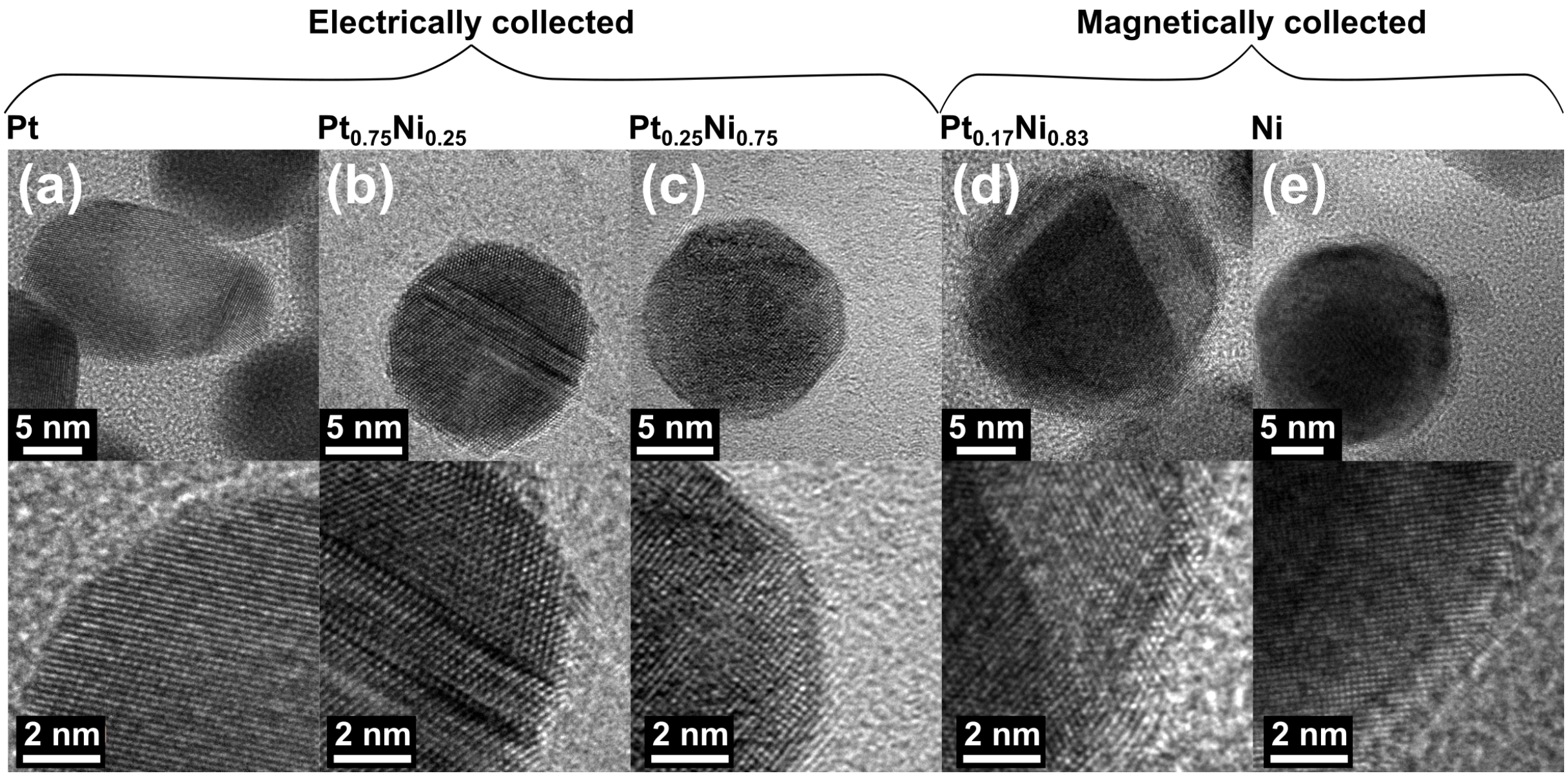

**Figure 3.** HRTEM images of the deposited PtNi nanoparticles. The particles typically consist of 1-3 different crystal domains. An oxide surface layer is only observed for the Ni-rich particles. For the Ni-rich nanoparticles there is no apparent separation into Ni rich and Pt rich domains.

The electrochemical performance towards HER of the nanotruss electrodes were tested and evaluated in 0.5 M $H_2SO_4$. **Figure 4**(a) shows linear sweep voltammograms for the $Pt_xNi_{1-x}$ samples with the composition specified in the figure label. In line with expectations the sample containing no Pt (x = 0) shows a very poor activity for HER, albeit with a strong improvement over time as displayed in the stability study shown in Figure S3(a). More noteworthy is that the nanotruss electrodes already at very low Pt-content (x = 0.03) reaches impressive values for the HER activity, and that the $Pt_{0.10}Ni_{0.90}$ nanotruss electrode initially outcompetes all other sample compositions including the Pt-reference sample, requiring only a potential of 17 mV to reach 10 mA cm$^{-2}$. This is comparable or better than most other reported data on HER electrocatalysts.[5, 27-29, 56-58]

The Tafel plot in Figure 4(b) shows that all $Pt_xNi_{1-x}$ nanotruss electrodes feature a Tafel slope of 30 mV dec$^{-1}$ at low current densities, in agreement with the Pt reference electrode, which implies very fast reaction kinetics and high conductivity and that the chemical recombination desorption of $H_2$ (Tafel reaction) is the rate determining reaction step.[48, 59] This further shows that the reaction mechanism of the $Ni_xPt_{1-x}$ nanotruss electrodes resembles that of the pure Pt electrode and that the surface of the $Ni_xPt_{1-x}$ nanoparticles on the nanotruss electrode probably are rich in Pt atoms, even for the Pt-dilute $Ni_{0.97}Pt_{0.03}$ sample. However, the linear Tafel regions at low current densities are relatively narrow in all $Ni_xPt_{1-x}$ electrodes, which could result from that there are several reaction sites with similar activity but with different hydrogen adsorption energies ($\Delta G_{\mathrm{H}}$).[48] In particularly, positive and negative $\Delta G_{\mathrm{H}}$ with similar magnitude would result in a combined reaction mechanism with a mixed Tafel slope. A non-existent linear Tafel slope could also be derived from negative $\Delta G_{\mathrm{H}}$, but still close to optimal ($\Delta G_{\mathrm{H}} \sim 0$ eV).[48] Thus, it is highly likely that the $Ni_xPt_{1-x}$ alloys do not only contain one type of active reaction site as in pure Pt, but have a more complex mechanism. We discuss and test this further below. In all samples, the Tafel slopes increase to 120 mV dec$^{-1}$ at higher current densities, indicating a stabilized hydrogen coverage, which is generally true for any metallic catalysts. Since the linear Tafel regions are relatively narrow it is difficult to extrapolate exchange current densities ($i_0$) with high accuracy. Nevertheless, in the low current density region between 5-30 mA cm$^{-2}$ the Tafel slopes are most linear, and $i_0$ of $Ni_{0.90}Pt_{0.10}$ and $Ni_{0.95}Pt_{0.05}$ are then calculated to -2.6 mA cm$^{-2}$ and -1.8 mA cm$^{-2}$, respectively.

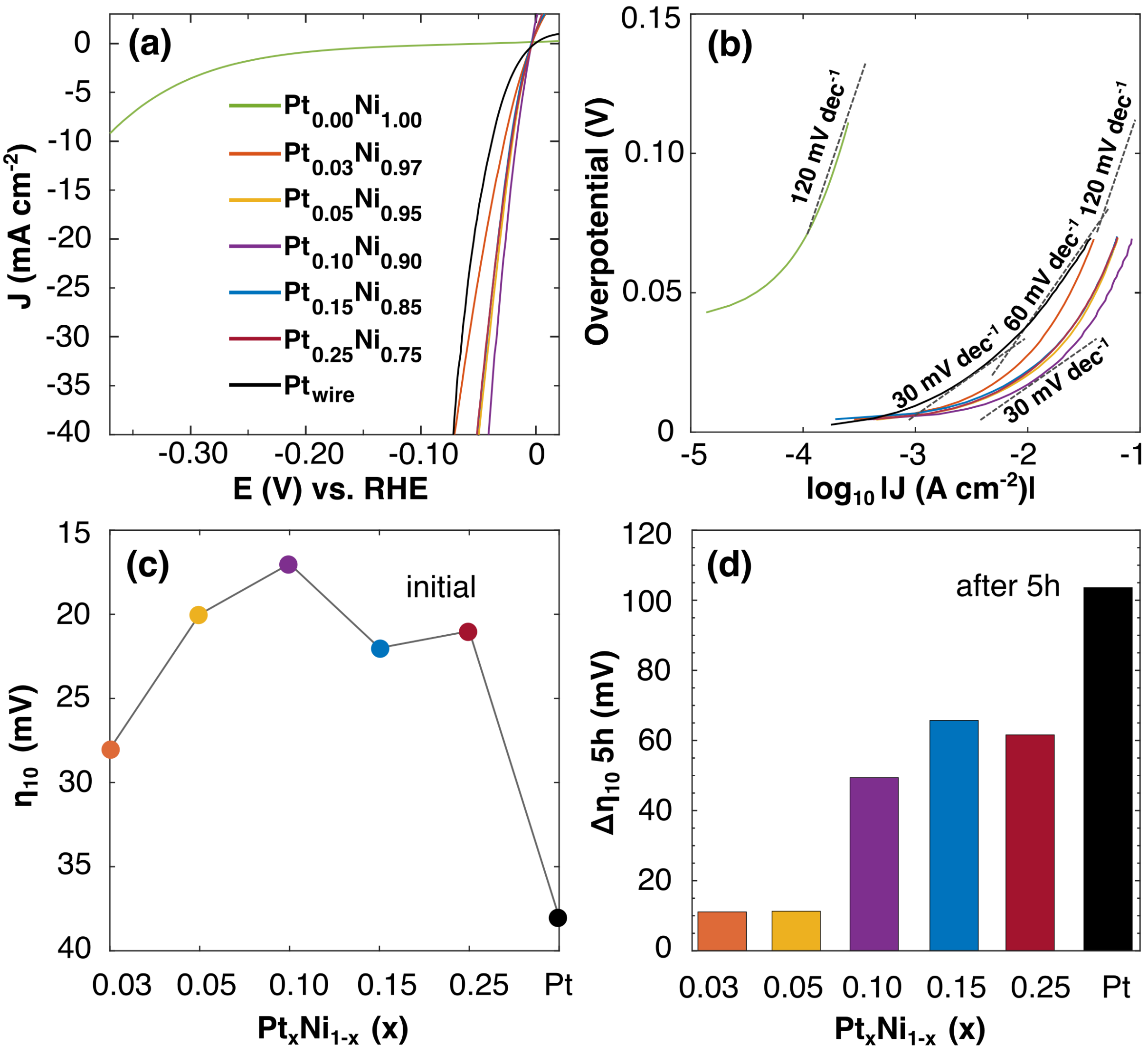

**Figure 4.** Electrochemical measurements. (a) Measured linear sweep voltammetry in 0.5 M $H_2SO_4$ for samples with various composition of $Pt_xNi_{1-x}$ and in (b) the corresponding Tafel plots with Tafel slopes (dashed) of 30 mV dec$^{-1}$, 60 mV dec$^{-1}$ and 120 mV dec$^{-1}$ as guides. (c) Initial overpotential for reaching a current density of 10 mA cm$^{-2}$ and (d) the additional overpotential needed at the same current density but after 5 hours tests at 10 mA cm$^{-2}$.

Figure S3 shows a measurement of the stability of the low-noble metal content $Pt_xNi_{1-x}$ nanotruss electrodes, as performed by registering the overpotential at 10 mA cm$^{-2}$ for 5 hours. The initial overpotentials for reaching 10 mA cm$^{-2}$ of the $Pt_xNi_{1-x}$ alloys are shown in Figure 4(c), and in Figure 4(d) the additional overpotential needed for reaching the same current density after the stability test is displayed. We find that the $Pt_{0.03}Ni_{0.97}$ and $Pt_{0.05}Ni_{0.95}$ nanotruss electrodes displays the best stability, as evidenced by that the required potential has increased by only 11 mV after five hours. It is also clear the in the $H_2SO_4$ electrolyte, the stability is better for the low

Pt content nanotruss electrodes, and that higher atomic ratio of Pt lead to faster drop in performance. Such phenomena are not seen in the $HClO_4$ electrolyte, as displayed in Figure S3 (b). This suggests that the Ni-rich nanotruss electrodes has a good ability to cope with adsorption of sulphur compounds from the electrolyte, which is known to cause problems for Pt-surfaces in process gas with high sulphur impurities.[60] The high stability of the nanotruss electrode is further verified by that the catalyst material remains firmly adhered to the carbon paper without any binder material during intensive bubble formation.

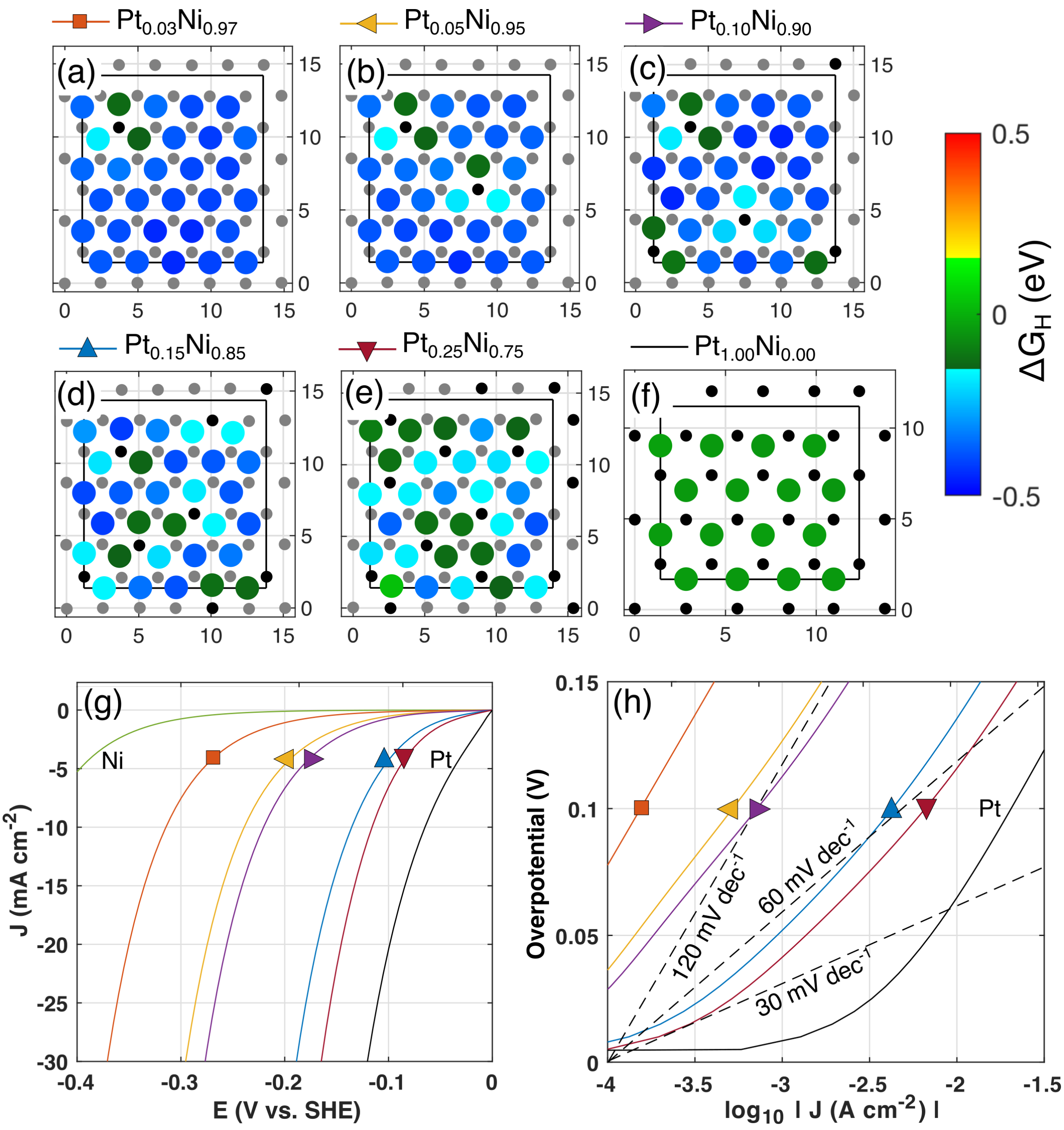

**Figure 5.** Theoretical results of $Pt_xNi_{1-x}$ systems. (a-e) Activity maps of $Pt_xNi_{1-x}(111)$ surfaces showing the final coordinate of adsorbed hydrogen for each reaction site as colored circles with the equilibrium values of $\Delta G_{\mathrm{H}}$ according to the color bar. The simulation cells are displayed as black lines and Ni and Pt atoms as gray and black circles, respectively. (f) Theoretical polarization curves for samples with various composition of $Pt_xNi_{1-x}$, and in (g) the corresponding Tafel plots with the dashed lines as guides.

To gain further insights to the high catalytic activity we performed ab initio calculations of $Pt_xNi_{1-x}$ alloys with density functional theory, and the results are displayed in **Figure 5**. To evaluate the alloys, we examined both 111 and 100 surfaces in the fcc lattice. In Figure 5(a-f) the 111 surfaces are shown, where coloured circles indicate the final positions of hydrogen and the equilibrium values of $\Delta G_{\mathrm{H}}$, considering self-consistency between the DFT calculated values of $\Delta G_{\mathrm{H}}$ and the hydrogen coverage according the Langmuir adsorption isotherm of the full surface.[48] The results from the 100 surfaces are shown in Figure (S4) and plots for the different sites displayed in another fashion are shown in Figure (S5). From Figure 5(a) it is clear that for each added Pt atom in the Ni(111) surface, three neighbouring adsorption sites are activated, displaying more optimal adsorption energies, leading to a high yield in activity gain for each added Pt atom. Theoretical voltammetry plots are further constructed by combining all reaction sites in the $Pt_xNi_{1-x}$ alloys that also comprise a non-interacting mix between 111 and 100-oriented crystal systems with a ratio of 1:1 using a recently developed microkinetic model.[48] From the voltammetry plots in Figure 5(f) it is clear that high performances comparable to the experimental samples are only reached for systems with the highest Pt content, i.e., $Pt_{0.15}Ni_{0.85}$ and $Pt_{0.25}Ni_{0.75}$. This indicates that all experimental alloys could have a high Pt content in the uppermost surface layer. On the other hand, the theoretical results represent flat surfaces with $1.5*10^{15}$ reaction sites $cm^{-2}$, and the high experimental current densities could possibly be due to high surface area of the 3D structures. Here we note however that the experimental and theoretical results of pure Ni agree well while having the same structure. Furthermore, the theoretical and experimental Tafel slopes should not be affected by surface area and can be directly compared. In Figure 5(g) we see that the theoretical Tafel plots also does not have any distinct linear regions of 30 mV $dec^{-1}$, which is caused by various reaction sites and overall slightly negative values of $\Delta G_{\mathrm{H}}$ The magnitude of the current densities could be more precisely compared if the electrochemical surface area (ECSA) of the experimental samples were known. However, we were not able to determine the ECSA, or catalyst loading accurately enough to make a fair comparison.

For theoretical calculations, a common issue for Pt-based materials is that theoretical $\Delta G_{\mathrm{H}}$ are slightly negative within the CHE model and Tafel slopes of 30 mV dec$^{-1}$ could thus not be obtained.[48, 61] Furthermore, the reaction mechanism on Pt is complex and it is possible that at during electrode polarization, hydrogen adsorbs on less preferable reaction sites for oversaturated surfaces leading to higher than one monolayer of H-coverage. Such additional hydrogen generally has more positive values of $\Delta G_{\mathrm{H}}$ that could result in Tafel slopes of 30 mV dec$^{-1}$. Yet, the trends and obtained exchange current densities from our results are still within high accuracy.

Our energetical studies (Figure S6) manifest that the Pt atoms prefer to be close to the surface rather than in the bulk, owing for a rich Pt surface for all $Pt_xNi_{1-x}$ alloys. We also observe that the surfaces are more stable for solid solutions where all Pt atoms are dispersed homogeneously rather than in cluster formation. Together, these two considerations will likely result in solid alloys with slightly Pt-rich surfaces, still avoiding clustering of Pt. Figure S6 shows simulated XRD patterns for a $Pt_{0.95}Ni_{0.05}$ with a diameter of 20 nm, where 15 % of the Pt-atoms have been segregated to the outer four layers, in line with the schematic Figure S7. The simulated XRD are in very good agreement with the Vegard's law and the experimental data (Figure 1(a)) and supports that a Pt-rich skin with fairly high Pt-content can form at the surface also for nanoparticles with very low Pt-concentration (Table S2). The HRTEM images in Figure 5 do not show any separate crystalline domains near the nanoparticle surface for the Ni rich particles. In summary the two last observations, show that the HER activity most likely originate from a partial Pt-segregation in the particles, which are incorporated into the Ni matrix. Other explanations such as a de-alloying and Ni-leaching from top atomic layers during electrochemical testing cannot be fully ruled out, but is less likely considering the initial high performance of the alloys and the increase of performance of the pure Ni sample over time in the chronopotentiometry measurements.

## 4. Conclusions

To conclude, the nanotruss $Pt_xNi_{1-x}$ electrocatalysts comprise a homogeneous isotropic metal bulk alloy, with a Pt enriched surface. The particles demonstrate a good conductivity, high surface area, and are strongly interlinked to each other as well as strongly adhered to the substrate. This leads to very strong HER activity. The best performing $Pt_{0.05}Ni_{0.95}$ sample demonstrates a Tafel slope of 30 mV $dec^{-1}$ in 0.5 M $H_2SO_4$ and an overpotential of 20 mV to reach 10 mA $cm^{-2}$. *Ab initio* calculations and simulations of X-ray diffraction data further show that the catalytic performance, and the stability in acidic medium, can be explained by the nanoparticle surface is Pt-rich. We also showed that each surface Pt atom in the Ni lattice will activate all neighbouring reaction sites, resulting in a high activity relative to Pt content. We believe that our results could be significant not only for electrolysis applications but also for other nearby research fields, such as fuel cells.

**Supporting Information**

Supporting Information is available from the Wiley Online Library or from the author.

**Acknowledgements**

T. W acknowledges support from Vetenskapsrådet (2017-04862) and Energimyndigheten (45419-1), and L. E. acknowledges support from Vetenskapsrådet (2017-04380) and Energimyndigheten (50779-1)

**Magnetically collected Platinum Nickel alloyed nanoparticles – insight into low noble metal content catalysts for hydrogen evolution reaction**

# Supporting information

Sebastian Ekeroth[1], Joakim Ekspong[2], Sachin Sharma[1], Robert Boyd[1], Nils Brenning[1, 3], Eduardo Gracia-Espino[b,] Ludvig Edman[2], Ulf Helmersson[1]*, Thomas Wågberg[2]*

[1)] Department of Physics, Linköping University, SE-581 83 Linköping, Sweden.

[2)] Department of Physics, Umeå University, SE-901 87 Umeå, Sweden.

[3)] KTH Royal Institute of Technology, School of Electrical Engineering, Division of Space and Plasma Physics, SE-100 44 Stockholm, Sweden.

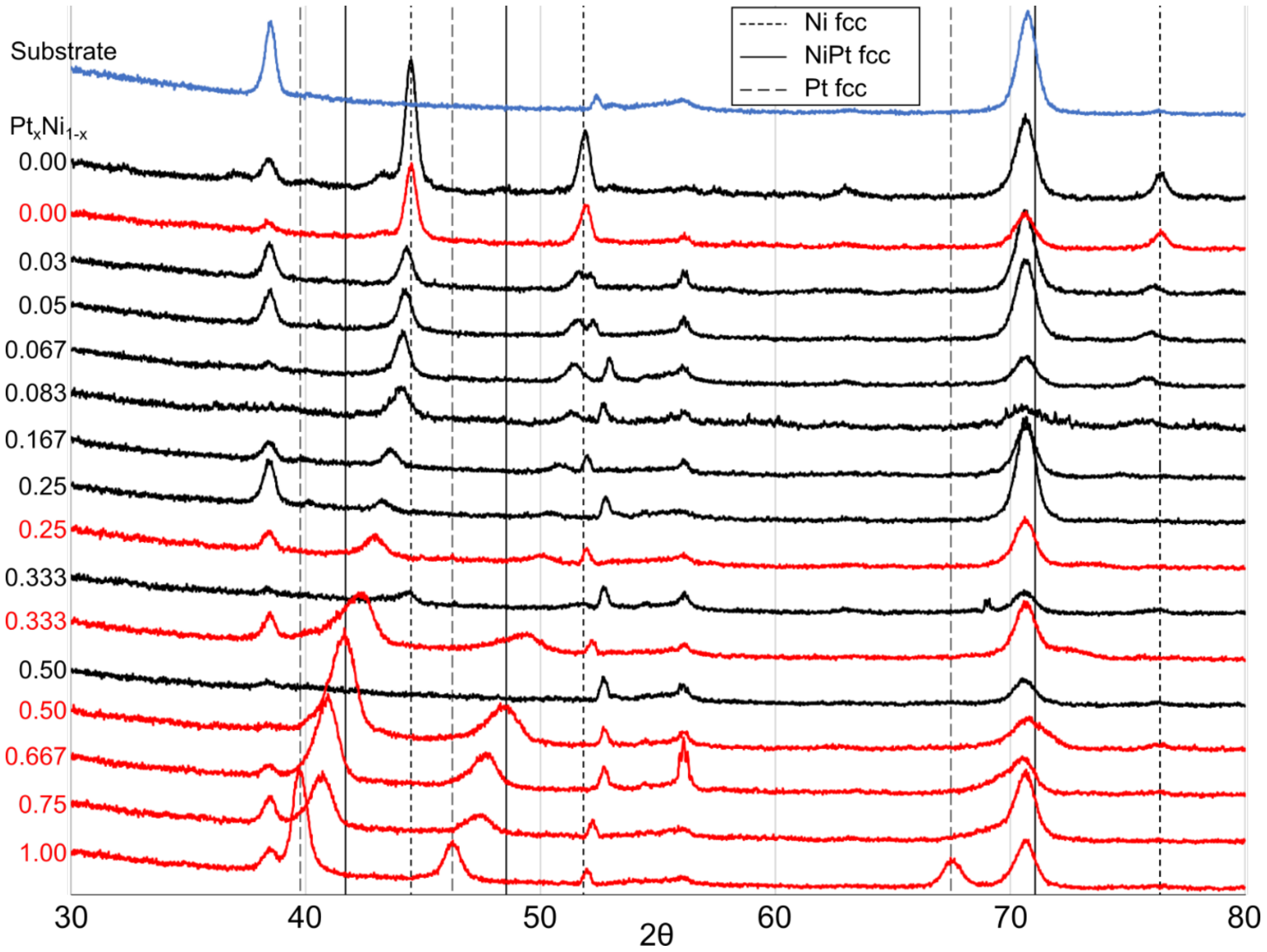

**Figure S1.** XRD of full gracing incidence scan 30-80°. Also displaying the scan of a substrate (Si wafer with 200 nm Ti). Magnetically collected nanoparticles are depicted in black, and electrically collected are in depicted in red. The nominal composition $Pt_xNi_{1-x}$ is given at the left of the curves.

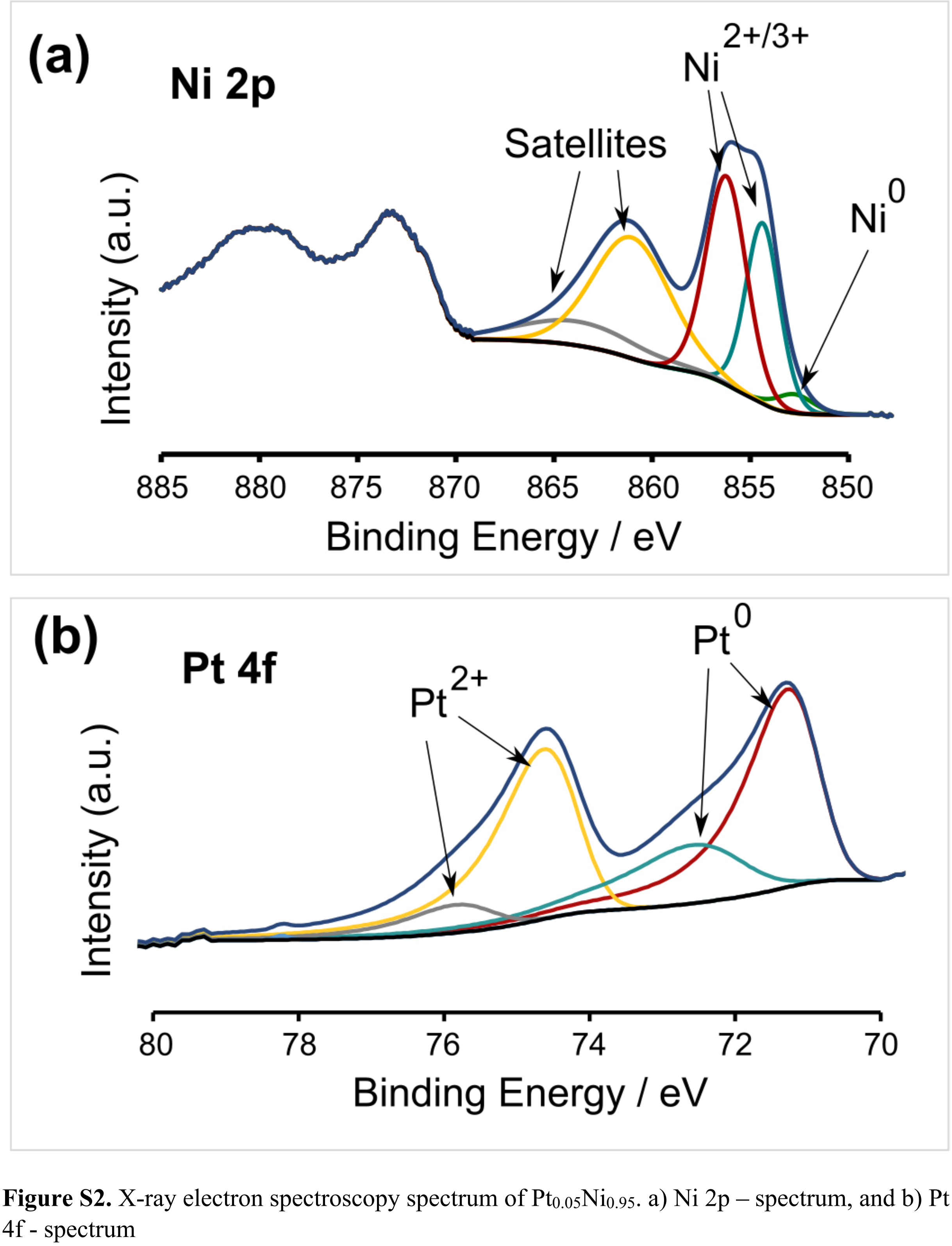

**Figure S2.** X-ray electron spectroscopy spectrum of $Pt_{0.05}Ni_{0.95}$. a) Ni 2p – spectrum, and b) Pt 4f - spectrum

**Table S1.** Process conditions and composition ratio for the different compositions.

| Nominal composition: $Pt_xNi_{1-x}$ x = | Power: Ni Cathode (W) | Power: Pt Cathode (W) | Total power (W) | Deposition time (min) | Collection method (Magnetic, Electric) | Catalysis measurements performed |
|---|---|---|---|---|---|---|
| 0.00 | 60 | 0 | 60 | 5 | M, E | X |
| 0.03 | 97 | 3 | 100 | 3 | M | X |
| 0.05 | 57 | 3 | 60 | 5 | M | X |
| 0.067 | 56 | 4 | 60 | 5 | M | |
| 0.083 | 55 | 5 | 60 | 5 | M | |
| 0.10 | 54 | 6 | 60 | 5 | M | X |
| 0.15 | 51 | 9 | 60 | 5 | M | X |
| 0.167 | 50 | 10 | 60 | 5 | M | |
| 0.25 | 45 | 15 | 60 | 5 | M, E | X |
| 0.333 | 40 | 20 | 60 | 5 | M, E | |
| 0.50 | 30 | 30 | 60 | 5 | M, E | |
| 0.667 | 20 | 40 | 60 | 5 | E | |
| 0.75 | 15 | 45 | 60 | 5 | E | |
| 1.00 | 0 | 60 | 60 | 5 | E | |

**Table S2.** Data for the Pt-nanoparticles where Pt-atoms have been segregated to the outer 4 layers at least 50 % of the surface with 15 % Pt-atoms

| **Average Pt content (at %)** | **Average Pt at the top layers (at %)*** | **Remaining Pt content (at %)**** |
|---|---|---|
| 0.0 | 0.0 | 0.0 |
| 3.0 | 7.3 | 2.3 |
| 5.0 | 7.3 | 4.6 |
| 6.7 | 7.3 | 6.6 |
| 8.3 | 7.3 | 8.5 |
| 10.0 | 7.3 | 10.4 |

* 15 at% Pt in sections along the surface covering 50% of it.

** At values above 6.7 at% of Pt, the remaining Pt content is higher than the average because of the restrictions impose on the Pt content in the top layers.

**Table S3.** Peak position, lattice constant and calculated compositions of the XRD peaks in Fig. S1. The peak positions have been determined using the "Peak parameters" tool in the software "Data Viewer".

| Nominal composition | Collection method | Peak position (2θ) [°] | Lattice constant (Bragg's law) [Å] | Calculated composition (Vegard's law) |
|---|---|---|---|---|
| 0.00 | M | 44.462 | 3.5251 | 0.0026 |
| 0.00 | E | 44.71 | 3.5244 | 0.0010 |
| 0.03 | M | 44.27 | 3.5396 | 0.0387 |
| 0.05 | M | 44.219 | 3.5435 | 0.0483 |
| 0.083 | M | 44.088 | 3.5535 | 0.0731 |
| 0.167 | M | 43.584 | 3.5925 | 0.1700 |
| 0.25 | M | 43.227 | 3.6207 | 0.2401 |
| 0.25 | E | 43.007 | 3.6384 | 0.2838 |
| 0.333 | M | N/A | N/A | N/A |
| 0.333 | E | 42.335 | 3.6934 | 0.4204 |
| 0.50 | M | N/A | N/A | N/A |
| 0.50 | E | 41.61 | 3.7548 | 0.5728 |
| 0.667 | E | 40.913 | 3.8160 | 0.7246 |
| 0.75 | E | 40.739 | 3.8316 | 0.7633 |
| 1.00 | E | 39.727 | 3.9251 | 0.9953 |

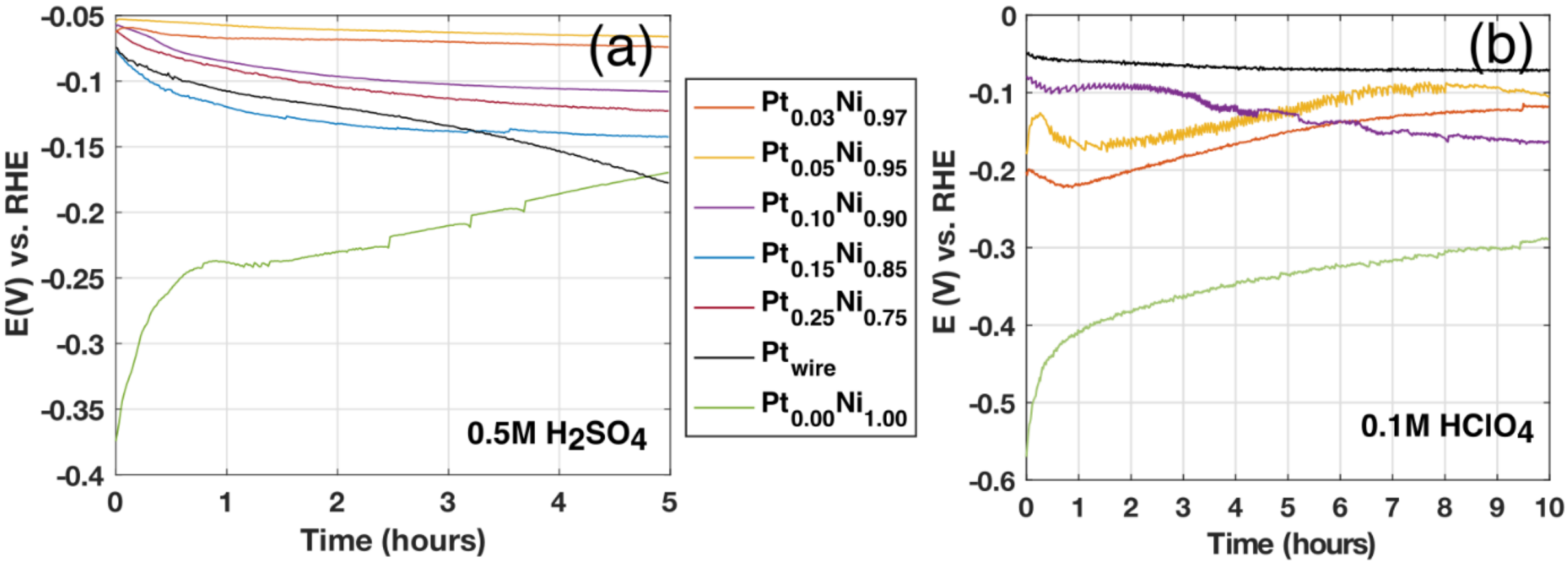

**Figure S3.** Stability tests. Stability tests for the samples over 5 hours of a constant current of 10 mA/cm$^2$ in (a) 0.5 M $H_2SO_4$., and (b) 0.1 M $HClO_4$.

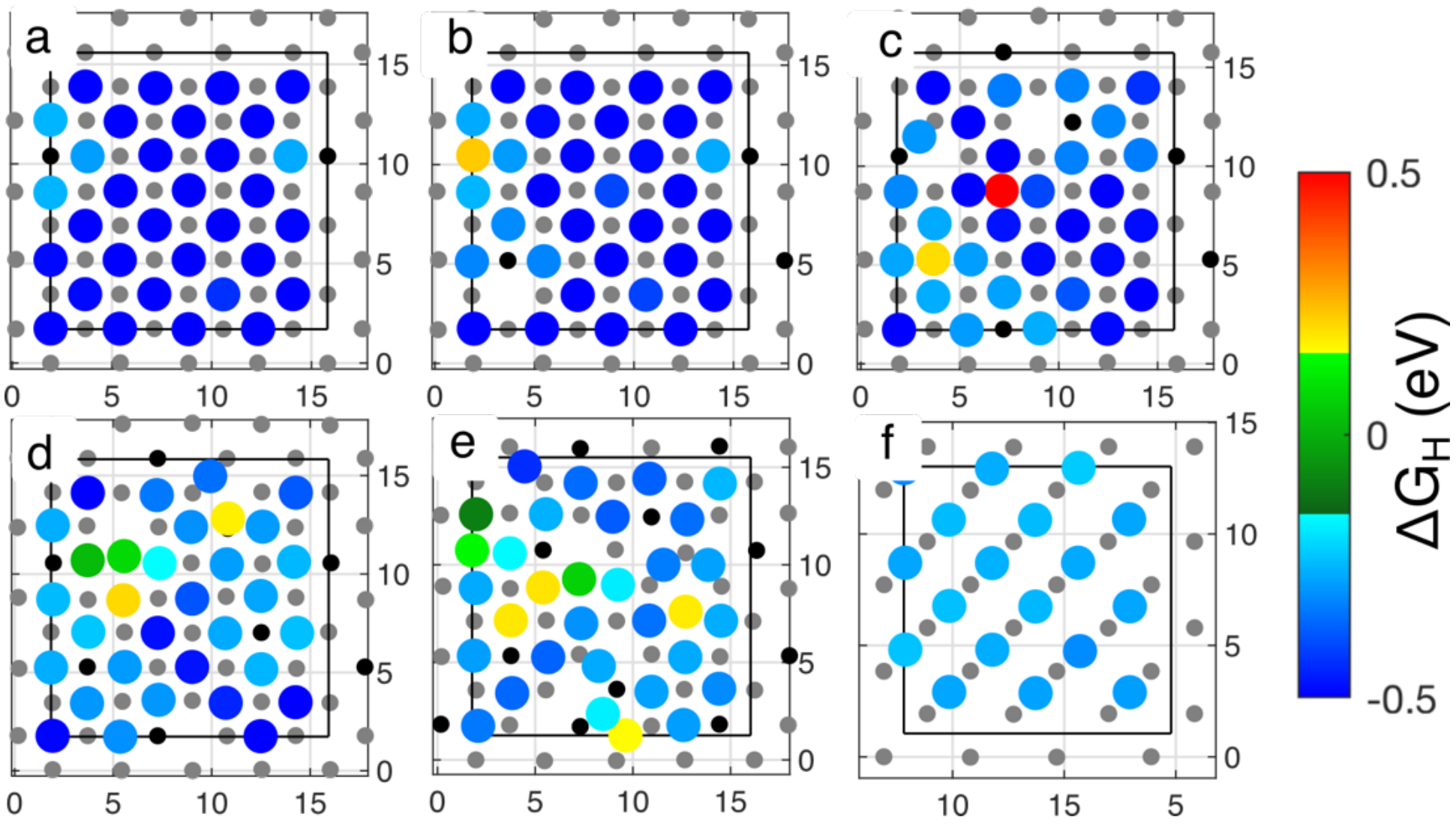

**Figure S4.** Activity maps. $PtNi_{100}$ alloys in (a-e) together with pure $Pt_{100}$ in (f). The larger colored circles show the optimized position and $\Delta G_H$, where optimal values are displayed as green. The grey and black circles marks nickel and platinum atoms, respectively.

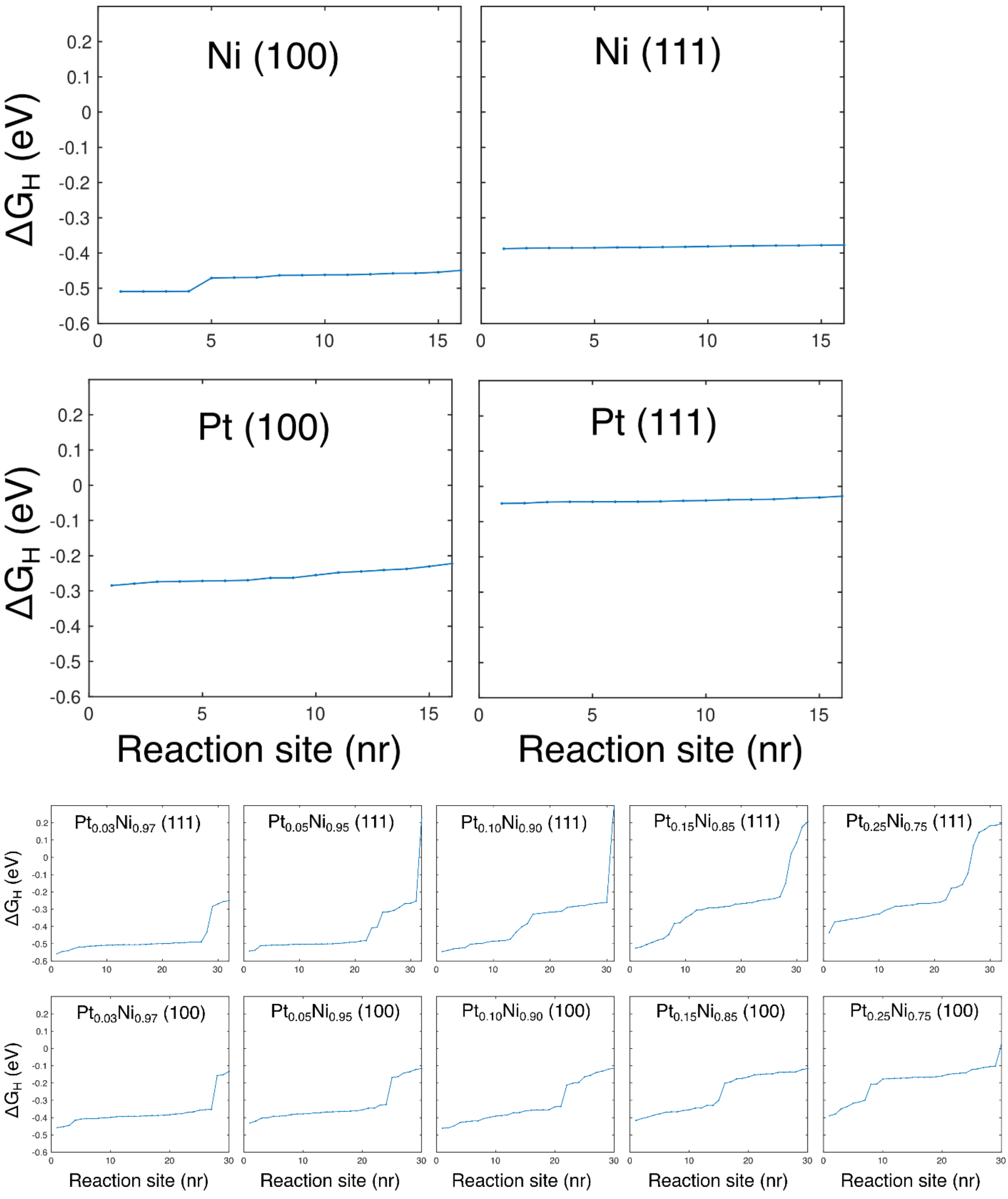

**Figure S5.** Equilibrium values of $\Delta G_{\mathrm{H}}$ for the specific sites in pure Pt and Ni with indicated crystal orientation together with the sites in PtNi-alloys in the indicated crystal orientation. For clarity, the data is sorted with respect to $\Delta G_{\mathrm{H}}$.

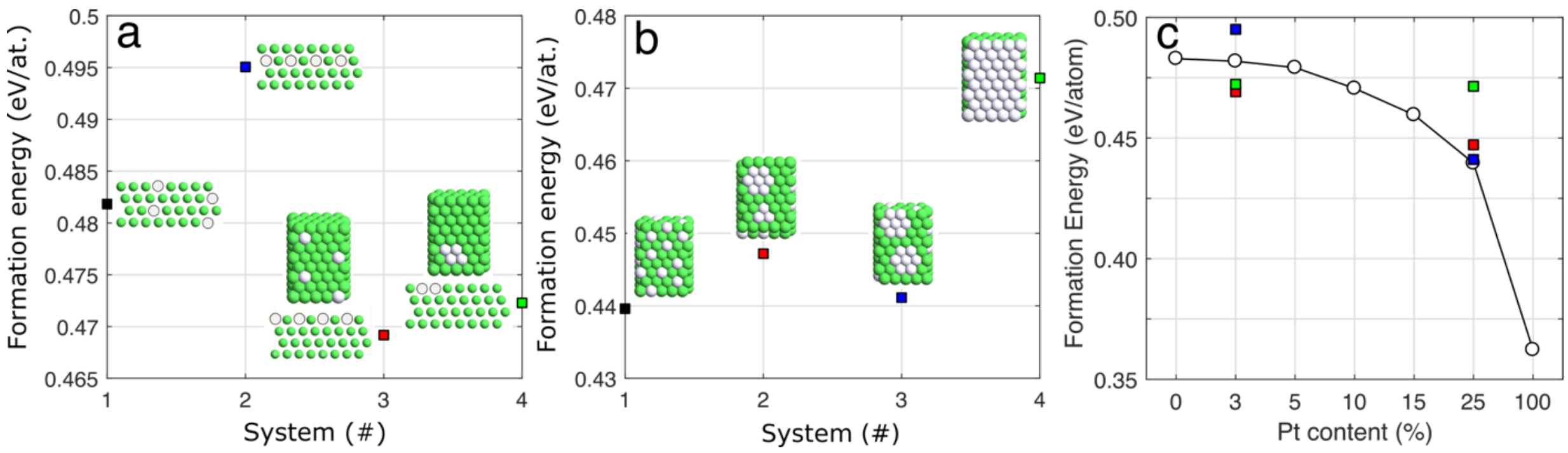

**Figure S6.** Calculated formation energies for different systems of the $NiPt_{111}$ alloys. In (a-b), the green circles correspond to nickel atoms while white circles correspond to platinum atoms. The top and side views are illustrated in (a) while only top views are shown in (b). In (c), the formation energy for $NiPt_{111}$ alloys used are shown with different Pt content. For 3 and 25 % Pt, the formation energies from (a-b) are also shown with matching colors.

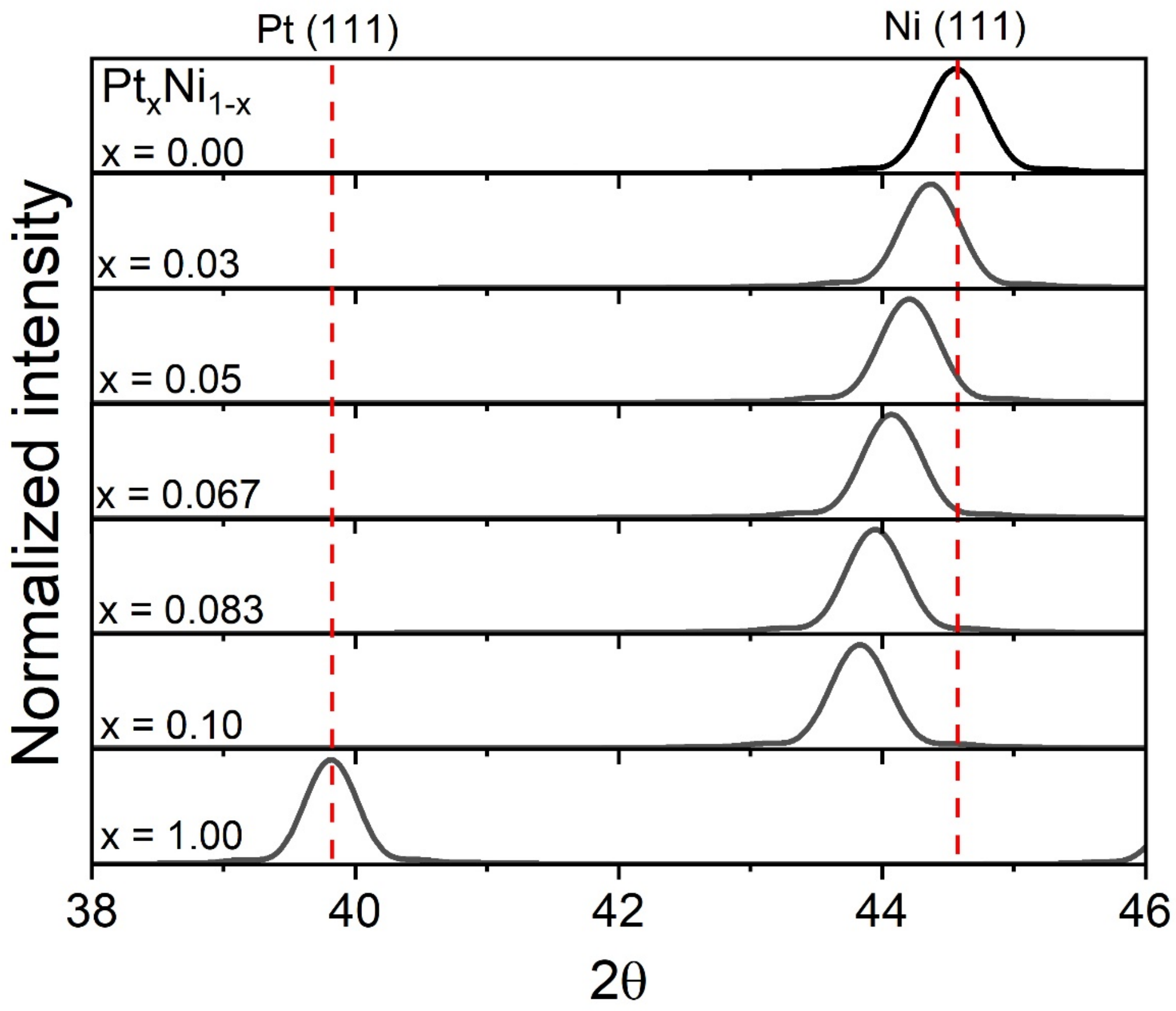

**Figure S6.** Simulated powder XRD of PtxNi1-x nanoparticles with an average size of 20 nm. The particles exhibit the average Pt content stated in the plot, however a larger localized Pt content of 15 at% was impose on all of them, as exemplified in Figure 1. Under this configuration, it is still possible to observe the displacement of the (111) peak according to Vegard's law and in agreement with the experimental data.

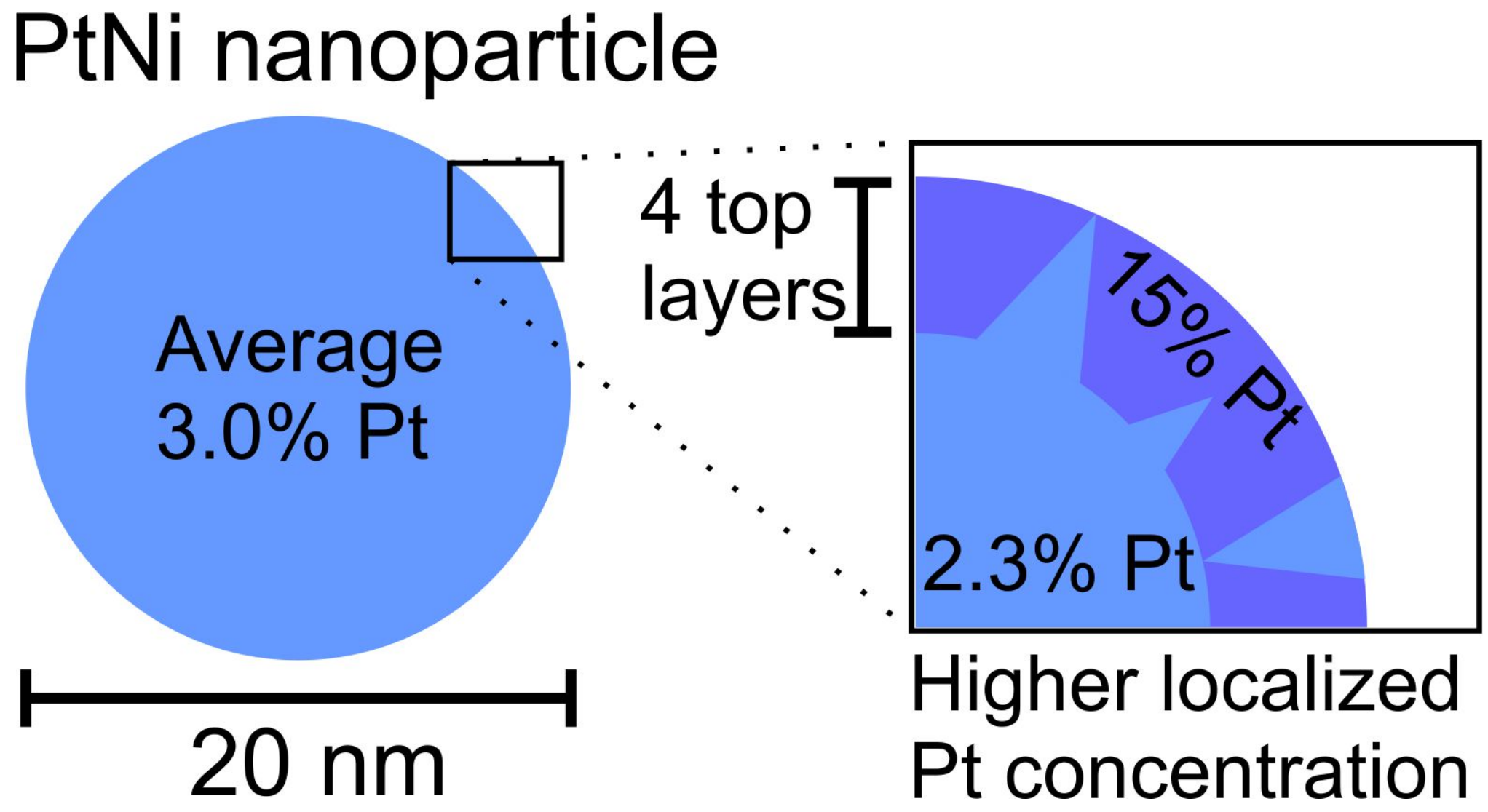

**Figure S7**. Schematic representation of a PtNi nanoparticle with an average Pt concentration of 3 at%. It is feasible to have a higher localized Pt concentration of 15 at% in the outer 4 atomic layers (~5 Å) covering at least half the surface. In other words, a complete 15 at% Pt shell is not possible, but several Pt rich sections with 15 at% covering at least 50% of the surfaces are plausible. Similar scenarios can be found for particles containing 5, 6.7, 8.3, and 10 at% of Pt, see table S2.